\documentstyle[preprint,pra,aps]{revtex}
\begin{document}
\tightenlines
\draft
\title{\bf
Enhancement of parity violating mixing
in halo nuclei and the problem of neutron weak parity 
nonconserving
potential constant
}
\author{M.S. Hussein$^{\dagger}$, A.F.R. de Toledo Piza$^{\dagger}$, 
O.K. Vorov$^{\dagger}$
and A.K. Kerman$^*$ }
\address{
$^{\dagger}$ Instituto de Fisica, \\
Universidade de Sao Paulo ,  \\
Caixa Postal 66318, 05315-970,  \\
Sao Paulo, SP, Brasil \\  
$^*$ Center for Theoretical Physics, \\
Laboratory for Nuclear Science and Department of Physics,\\
Massachusetts Institute of Technology,\\
Cambridge, Massachusetts 02139, USA
}
\date{15 Yuly 1999}
\maketitle
\begin{abstract}
We consider the Parity Nonconserving (PNC) 
mixing in the ground state of 
exotic (halo) nuclei caused by the 
PNC weak interaction 
between outer neutron and nucleons {\it within nuclear interior}.
For the nucleus $^{11}Be$ as an example of typical nucleus with 
neutron halo, we use analytical model for the external neutron 
wave functions to estimate the scale of the PNC mixing.
The 
amplitude
of the PNC mixing in halo state 
is found 
to be 
an order 
of magnitude bigger
than that of typical PNC mixing between the ``normal'' nuclear 
states in nearby nuclei.
The enhanced PNC mixing in halo cloud is proportional to the 
neutron weak PNC potential constant $g^W_n$ only.  
\end{abstract}

\section{Introduction}
\label{sec:level1}

The parity nonconserving (PNC) nucleon interaction in nuclei
caused by the PNC Weak interaction,
and PNC effects in neutron-nucleus reactions
are subject of current interest for both experimentalists and 
theorists 
\cite{DDH,ADELBERGER,te0,te1,pst,DESP,WB,KSSW,FHKV}.
The overall scale of the observable PNC effects is found to be 
in reasonable 
agreement with estimates in existing theory of the weak 
interactions 
\cite{DDH}
based on the Standard Model. 
Complete understanding of PNC forces in nuclear domain, which
requires reliable QCD-based models of hadrons is 
far from being reached. 
This motivates extensive studies of the strengths of the PNC forces.  

So far, the PNC effects have been probed in ``normal'' nuclei.
Physics of ``exotic'' nuclei studied with unstable nuclear beams 
\cite{ALK-TOST,JO-ALK-TO,SAGAWA,HALO-OBSOR,HU-PLB,Hencken-Bertsch-Esbensen,PAPER-RESTORATION,HALO-WEAK,HALO-BETA,Brown-Hansen,Karataglidis-Bennhold,Ren-Faessler-Bobyk,SY}
appears to be one of the most 
promising
modern nuclear areas.
Due to their specific structure, exotic nuclei, e.g., halo nuclei 
can 
offer
new possibilities to probe those aspects of nuclear interactions 
which are not accessible with normal nuclei.  
It is therefore interesting to examine possibilities of 
using exotic nuclei to investigate the effects of violation of
fundamental symmetries, i.e., spatial parity and time reversal.

Some aspects of the Weak interactions in exotic nuclei have 
been discussed in 
literature \cite{HALO-WEAK},\cite{HALO-BETA} in relation to the 
beta decay and to possibilities to study the parameters of the
Cabibbo-Kobayashi-Mascawa matrix.
To the best of our knowledge, however, the issue of the PNC effects
in exotic nuclei has not been addressed yet. 

The aim of this work is to present a first 
evaluation of the 
magnitude of the
PNC effects 
in halo nuclei. 
We confine ourselves to the case of nucleus $^{11}Be$, 
the most well studied,
both experimentally and theoretically 
\cite{ALK-TOST,JO-ALK-TO,SAGAWA},
\cite{Hencken-Bertsch-Esbensen},\cite{PAPER-RESTORATION}  . 
We find that the ground state, the $2s_{1/2}$ halo configuration, 
acquires
admixture of the closest in energy halo state of opposite
parity,
$1p_{1/2}$. 
This effect originates from the weak interaction of the external 
halo
neutron with the core nucleons in the nuclear interior.
As a result, the neutron halo cloud surrounding the nucleus 
acquires the wrong parity admixtures that may be tested 
in experiments which can probe the halo wave functions in the
exterior. 

The magnitude of the admixture is found to be 
$\sim 10^{-6} \times g^W_n$ that is an order of magnitude 
bigger than the PNC effects in normal spherical nuclei. 
What is important to notice is that 
the enhanced effect we have found here
is proportional to the {\it neutron weak constant} $g^W_n$ only. 
The value of this constant remains to be one of the most 
questionable 
points in modern theory of parity violation in nuclear 
forces \cite{pst}.
The enhanced PNC mixing in halo found here can be therefore 
useful 
in studies of the neutron weak constant.

One should mention 
another interesting question related to the 
structure of the PNC force in nucleus, namely, the strength of 
the isovector 
P-odd potential that has been discussed in Refs.
\cite{DESP,WB}.

In the next two sections, we recall the basic facts about 
the PNC weak 
interaction between nucleons 
starting from
the Hamiltonian of Desplanques, Donoghue and Holstein
(DDH). We consider the potential approximation to 
relate the parameters of the initial PNC Hamiltonian to the 
single-particle PNC mixing effects in nuclei, 
and discuss the problem of the neutron PNC constant in Sec.III.
In Sec.IV, we analyze the basic effects of the halo structure 
on the magnitude of the single-particle PNC mixing and make estimates. 
In Sec.V, we use analytical approximation for the halo 
wave functions in $^{11}Be$ to calculate the matrix elements 
of the PNC weak interaction involving halo states and to 
estimate the magnitude of PNC mixing in the ground state.
Sec.VI summarizes the results and presents brief discussion of 
their implications.

\section{Weak nucleon-nucleon interaction and parity violating 
effects. Potential approximation}
\label{sec:level2}

We start with writing the nuclear Hamiltonian $H$ in the form
\begin{equation} \label{TOTAL-H}
H=H^{0}_S \quad + \quad V^{res}_{S} \quad + \quad W^{PNC} \quad ,
\end{equation}
where the first term 
$H^{0}_S=\sum_a(\vec{p_a}^{2}/2m+U_{S}(r_a))$ is the single particle 
Hamiltonian of the nucleons including the single-particle piece 
$U_{S}$ of 
the strong interaction, $V^{res}_{S}$ is the residual
two-body strong interaction.  The last term, $W^{PNC}$ 
is the PNC part 
of the Weak interaction that is the source of the PNC effects. 

The magnitude of the PNC effects is sensitive to both 
the weak PNC interaction 
matrix elements between the states of opposite parity and to 
the nuclear structure effects given by the strong part of the 
Hamiltonian (\ref{TOTAL-H}). The latter one
is invariant under
spatial coordinate reflections,
and if there is no weak interaction term $W^{PNC}$ 
in (\ref{TOTAL-H}),
and as such parity is preserved, the eigenstates $|\Psi_s \rangle$
of the strong Hamiltonian $H^0_S + V^{res}_S$ with energies $E_s$
can be labeled by the 
parity quantum number (positive or negative), 
$|\Psi^+_s \rangle$,  $|\Psi^-_s \rangle$.
Due to presense of PNC weak interaction $W^{PNC}$ in the nuclear 
Hamiltonian (\ref{TOTAL-H}), a state of definite parity, say, 
$|\Psi^+_s \rangle$, 
acquires very small 
admixtures of wrong parity configurations. This can be accounted for
by using the first order of perturbation theory with respect 
to $W^{PNC}$:
\begin{equation} \label{PERTURBATION}
| \Psi^+_s \rangle' = | \Psi^+_s \rangle +
\sum_{ s1} \frac{ \langle  \Psi^-_{s1} | W^{PNC} | \Psi^+_s \rangle}
{ E_{s} - E_{s1} } | \Psi^-_{s1} \rangle.  
\end{equation}
Here, prime denotes the corrected wave function that accounts for the 
PNC interaction and sum goes over available states of opposite parity,
$| \Psi^-_{s1} \rangle$.
The magnitude of measurable PNC effects is normally 
proportional to the 
coefficients $f^{PNC}$ \cite{ADELBERGER}
that determine the dominating admixtures of the wrong parity states
\begin{equation}   \label{F-MIXING}
f = \frac{  \langle  \Psi^-_{s1} | W^{PNC} 
| \Psi^+_s \rangle}{ \Delta E}.
\end{equation}
The natural scale of the PNC effects in nuclei under usual
conditions is \cite{DDH},\cite{ADELBERGER}
\begin{equation} \label{F-SCALE}
| f | \simeq 10^{-7}
\end{equation}
that is roughly the ratio of the strength of the 
Weak PNC forces ( matrix element in 
the numerator of (\ref{F-MIXING}) ) and the strength of the 
strong interaction (energy denominator in (\ref{F-MIXING}) ).
In highly selective experiments,
the PNC effects can be enhanced
considerably as compared to estimate (\ref{F-SCALE}),
due to specific properties of a specially chosen nuclear 
system or process. 
To reach high sensitivity to the wrong parity admixtures,
one usually seeks possibilities to have 
the denominator $\Delta E$ in
(\ref{F-MIXING}) minimal while keeping the PNC matrix element 
at maximum and to  
improve selectivity of measurable effect.
This is typical for any tests of fundamental symmetries. 

The most widely used version of the microscopic 
PNC interaction is the 
DDH Hamiltonian $W^{PNC}_{DDH}$ \cite{DDH}, 
where the PNC forces are mediated by mesons.
Its form stems from the analysis of interactions 
between intranucleon
quarks via exchange of heavy bosons of Standard Model.
The nonrelativistic P-odd weak 
interaction between nucleons 
approximated by the one-meson exchange can be written in the 
form \cite{ADELBERGER},\cite{DDH}
\begin{eqnarray}  \label{DDH-WEAK}
W^{PNC}_{DDH} = 
i \frac{h^{(1)}_{\pi} g_{\pi}}{4 \sqrt{2}m}
(\vec{\tau}_1 \times \vec{\tau}_2)^{(3)} 
(\vec{\sigma}_1+\vec{\sigma}_2)
[\vec{p}_1-\vec{p}_2, 
{\cal F}_{\pi} ] -
\nonumber\\
- \frac{g_{\rho} h^0_{\rho}}{2 m}
(\vec{\tau}_1 \cdot \vec{\tau}_2) 
(\vec{\sigma}_1-\vec{\sigma}_2)
\{ \vec{p}_1-\vec{p}_2,  
{\cal F}_{\rho} \} -
\nonumber\\
- \frac{g_{\rho} h^0_{\rho}}{2 m} i (1+\mu)
(\vec{\tau}_1 \cdot \vec{\tau}_2)
(\vec{\sigma}_1 \times \vec{\sigma}_2)
[ \vec{p}_1-\vec{p}_2,  
{\cal F}_{\rho} \} 
+ W',
\end{eqnarray}
where the standard notations 
${\cal F}_{\pi(\rho)}=
\frac{e^{-m_{\pi(\rho)}|\vec{r}_1-\vec{r}_2|}}{4 \pi
|\vec{r}_1-\vec{r}_2|}$ are used and $[.,.]$ and $\{.,.\}$ denote the
commutator and anticommutator,
respectively. The subscripts $1$ and $2$ label the interacting
nucleons,
the superscript $(3)$ denotes the third isospin projection.
Here, $m$, $m_{\pi}$ and $m_{\rho}$ are the masses of the nucleon, 
$\pi$- and $\rho$-meson, respectively; $\sigma$ ($\tau$) stand for
the
spin (isospin) Pauli matrices, $\mu=3.7$ is the isovector part
of the anomalous magnetic moment of nucleon.
$W'$ denotes contributions from heavier mesons,
which are less important.
The values of the 
corresponding weak and strong coupling constants 
$h^{(1)}_{\pi}$, $g_{\pi}$, $g_{\rho}$, and $h^{0}_{\rho}$ 
can be found in \cite{DDH},\cite{ADELBERGER}. 

In nuclear environment, a nucleon experiences the combined action
of the PNC forces (\ref{DDH-WEAK}) from other nucleons. 
It is known, 
see, e.g., \cite{ADELBERGER}, 
that the most of P-odd effects caused by the
weak interaction $W^{PNC}_{DDH}$ (\ref{DDH-WEAK}) in (\ref{TOTAL-H}) 
can be successfully modeled by introducing 
the effective one-body P-odd
interaction, or the ``weak potential'', $W_{sp}$,
acting on the nucleon $1$ as a single-particle operator
which arises from
averaging $W^{PNC}$ over the states  of other nucleons 
$W_{sp} \equiv  \langle W^{PNC} \rangle$.
Within this approximation, the
Hamiltonian of the weak interaction 
in a nucleus
takes particularly simple form of a sum of the proton $W^p_{sp}$ and 
neutron $W^n_{sp}$ symmetry violating potentials
\begin{eqnarray}  \label{WEAK-POTENTIAL}
W_{sp}= W^{p}_{sp}+W^{n}_{sp} =
g^{W}_{p}\frac{G}{2\sqrt{2}m}
\lbrace({\vec \sigma}_{p}{\vec p}_{p}),\rho\rbrace 
+
g^{W}_{n}\frac{G}{2\sqrt{2}m}
\lbrace({\vec \sigma}_{n}{\vec p}_{n}),\rho\rbrace, 
\end{eqnarray}
where $G=10^{-5}m^{-2}$ is the Fermi constant, 
$\vec{p}_{p(n)}$ and $\vec{\sigma}_{p(n)}$ 
refer to the proton (neutron) momentum and doubled spin 
respectively.
The coherent contribution from all the 
occupied nucleon orbitals
composing the core 
yields the nuclear density 
$\rho = \sum_{occ} |\psi_{occ}|^2$ 
in the expression (\ref{WEAK-POTENTIAL}).
The dimensionless constants $g^W_p$ and $g^W_n$ of 
order of unity, for the 
proton and neutron potentials, are related to the parameters of the 
DDH Hamiltonian and depend on nuclear charge and neutron number. 
The single-particle approximation (\ref{WEAK-POTENTIAL})
for the PNC weak interaction (\ref{DDH-WEAK}) 
turns out to be very accurate \cite{ADELBERGER}. 
It works satisfactorily even in the case of compound nuclear states 
\cite{te1},\cite{we,dress,A-B}
where (\ref{WEAK-POTENTIAL}) gives the dominating 
contribution \cite{we},\cite{dress} despite the fact that the wave 
functions
are of essentially many-body nature.

\section{
Proton and neutron weak potential strengths
}
\label{sec:level3}

The knowledge
about the proton and neutron constants $g^W_p$ and $g^W_n$ 
accumulated
to date can be summarized as follows:
\begin{equation} \label{g-values}
g^W_p = 4.5 \pm 2 ,  \qquad g^W_n = 1 \pm 1.5.
\end{equation}
These widely used
values \cite{ADELBERGER,pst,FKh,we}
%
correspond to the best values \cite{DDH} of the
microscopic parameters in the DDH
Hamiltonian (\ref{DDH-WEAK})
and they are found in reasonable 
agreement with 
the bulk experimental data on parity violation,
including the recent compound nuclear experiments 
\cite{te0}
and
anapole moment measurements \cite{EXP-AM-SC}.
The above relatively small absolute value of the neutron 
constant that follows
from DDH analysis, results
basically from cancellation between $\pi$- and $\rho$-meson
contributions to  $g^W_n$, while both mesons contribute coherently 
to the proton constant $g^W_p$, see, e.g.\cite{pst}.  
Due to this difference between the 
absolute values of the proton and neutron constants,  
the proton constant tends
to dominate most measurable PNC effects
\cite{ZELD,FKh,HH-AM,BP-AM},
especially when both $g^W_p$ and $g^W_n$ can contribute.
In some cases (such as odd proton nuclei), 
the contribution from the neutron constant,
$g^W_n$, is suppressed irrespectively of its strength
\cite{ADELBERGER},\cite{DESP}.
In this sense, one usually measures the value of $g^W_p$, and
it is difficult to probe $g^W_n$ unless special suppression 
of the proton contribution occurs, and/or contribution of $g^W_n$
is highlighted. 
By contrast, the case we consider in this work is sensitive 
to the value of the neutron constant only. 

\section{
Halo structure effects on the PNC mixing
}
\label{sec:level4}

The basic specific properties of the halo nuclei are 
determined by the fact 
of existence of loosely bound nucleon in addition to the 
core composed by the rest of the nucleons \cite{HALO-OBSOR}
(we will be interested here 
in the most well studied case of neutron halo).     
The matter distribution is shown schematically in Fig.1 (part a).

In one-body halo
nuclei like $^{11}Be$, the ground state is particularly
simple: it can be 
represented as direct product of the single-particle wave function
of the external neutron,
$\psi_{halo}$, and the wave function of the core.
The residual interaction $V^{res}_{S}$ in (\ref{TOTAL-H})
can be neglected
as the many-body effects related to the core
excitations are generically weak in such nuclei \cite{CORE-EXC}.
The problem with the Hamiltonian (\ref{TOTAL-H}) is reduced to a 
single-particle problem for the external nucleon.
The PNC potential matrix element between the
ground state of halo nucleus and a state with opposite parity is 
\begin{eqnarray} \label{W-SP-HALO}
\langle \psi^+_{halo} | W^{PNC} | \psi^-_{halo} \rangle = 
g^{W}_{n}\frac{G}{2\sqrt{2}m} \langle \psi^+_{halo} | 
\lbrace({\vec \sigma}_{n}{\vec p}_{n}),\rho_c\rbrace 
| \psi^-_{halo} \rangle ,
\end{eqnarray}
where $\rho_{c}(r)$ is the core density.
Due to relatively heavy core 
for $A\simeq 10$,
difference between 
the center of mass coordinate and the 
center of core coordinate
can also be neglected.  

The effective potential that binds external neutron is rather
shallow 
yielding
small single-neutron separation energy,
and one can expect small energy spacing between the opposite parity
states. The PNC effects (\ref{F-MIXING}), (\ref{PERTURBATION})
can therefore be considerably magnified.
The spectrum of $^{11}Be$ is shown in Fig.1(b). To evaluate the
PNC mixing $f^{HALO}$ in the ground state of this nucleus, 
it is enough
to know the single-particle matrix element between the ground state
2s and the nearest opposite parity state 1p, and use their energy
separation that is known experimentally.
 
The second effect of halo is that  
the value of the matrix element of the
operator    (\ref{WEAK-POTENTIAL}) between the halo states can 
be dramatically reduced as compared to its value 
in the case of ``normal''
nuclear states.
The single-particle weak PNC potential (\ref{WEAK-POTENTIAL})
in (\ref{W-SP-HALO}) originates from the DDH Hamiltonian 
(\ref{DDH-WEAK})
which is two-body operator, this fact is hidden in
the nucleon density of the
core $\rho_{c}(r)$.
The external neutron spends most of its time 
away from the core region
where only it can experience the PNC potential created by the
rest of nucleons.
Indeed, the dominant contribution to the matrix element of 
(\ref{WEAK-POTENTIAL}) between the halo states in (\ref{W-SP-HALO})
must come from the regions where the three functions can overlap 
coherently:
$\psi^+_{halo}(r)$,  $\psi^-_{halo}(r)$ and the core density 
$\rho_{core}(r)$.
The latter one is essentially restricted by the region of nuclear 
interior,
$r < r_{c}$ thus reducing the effective volume of required 
interference
region to $ \frac{4}{3} \pi r_{c}^3$. 
Normalization condition 
implies 
that 
the extended wave function of the bound state halo 
$\psi^{\pm}_{halo}(r)$ must be considerably reduced in the volume
of coherent overlap $ \frac{4}{3} \pi r_{c}^3$. 
By contrast, in ``normal'' nuclei the radii of 
localization of the wave
functions with opposite parity that can be mixed by 
the weak interaction
coincide generically with the core radius $r_{c}$.
The resulting suppression for the PNC halo matrix element    
$\langle\psi^-_{halo}(r) | W_{sp} | \psi^+_{halo}(r) \rangle$
with respect to the matrix element for the normal nuclei 
can be extracted from the following simple estimate
\begin{eqnarray} \label{ESTIMATE}
\frac{\langle\psi^-_{halo} | W_{sp} | \psi^+_{halo} \rangle}
{\langle\psi^-_{normal} | W_{sp} | \psi^+_{normal} \rangle}
\sim \left( \frac{ r_{c}}{ r_{halo}} \right)^3 
\nonumber\\
\sim \left( \frac{  2 fm }{ 6 fm } \right)^3 
\sim \frac{1}{25} ... \frac{1}{30} 
\end{eqnarray}
where we have used the mean square radii of halos from Ref.
\cite{SAGAWA}.
This suppression factor can cancel out the effect of the small
energy separation (the denominator in Eq.(\ref{F-MIXING}))
and to suppress the PNC effects.
This simple estimate does not account for 
the structure of the halo wave functions which can be quite
substantial and may even lead to further 
suppression in the PNC mixing.
In the following, we present a detailed analysis of the related
effects. 
In particular, we find that the crude estimate (\ref{ESTIMATE})
turns out rather pessimistic.

\section{
Halo Model and
Evaluation of the PNC mixing in the ground state of $^{11}Be$
}
\label{sec:level5}

The form of the single-particle wave functions 
of halo states can be deduced from their basic properties
\cite{PAPER-RESTORATION} 
and their quantum numbers \cite{SAGAWA}. The results of the 
Hartree-Fock calculations which reproduce the main halo properties
(e.g., mean square radii) are also available \cite{SAGAWA}. 
We use the following ansatz for the model wave function of the $2s$ 
halo state:
\begin{equation} \label{ANSATZ-2s}
\psi_{2s} = R_{2s}(r) \Omega^{l=0}_{j=1/2,m}, 
\quad
R_{2s}(r) = C_0 (1- (r/a)^2 ) exp(- r/r_0)
\end{equation}
Here, $R_{2s}(r)$ is the radial part of the halo wave function
and 
$\Omega^{l=0}_{j=1/2,m}$ is the spherical spinor. 
As we can neglect 
the center of mass effect for the heavy ($A=10$) core,
the halo neutron 
coordinate $r$ in
$R_{2s}(r) = 
\frac{1}{r}\chi_{2s}(r)$
is reckoned from the center of nucleus. 
The constant $C_0$ is 
determined from the normalization condition, 
$\int\limits_0^{\infty} dr [\chi_{2s}(r)]^2 = 1$ (we choose the
radial wave functions to be real). 
We have 
\begin{equation} \label{C0}
C_0 = \frac{ 2^{3/2} a^2 }
{ r_0^{3/2} \sqrt{45r_0^4 + 2a^4 - 12 a^2r_0^2     } }   
\end{equation}

The parameters $r_0$ and the 
$a$ are the corresponding lengths to fit the density distributions
obtained in Ref.\cite{SAGAWA} and the mean square radius.
The value of $a$ is practically fixed to be $a = 2 fm$ what 
corresponds to the position of the node. Recently, 
the node position
have been restored from the analysis of the 
scattering process in work \cite{PAPER-RESTORATION}. 

For the wave function $\psi_{1p} = R_{1p}(r) \Omega^{l=1}_{j=1/2,m}$
of the excited state $1p$, the following simplest form of the radial 
wave function turns out to be adequate
\begin{equation} \label{ANSATZ-1p}
R_{1p}(r) = C_1 r exp(- r/r_1),
\end{equation}
where $C_1$ is the normalization constant 
$
C_1 = \frac{ 2}{\sqrt{3}} r_1^{-5/2}
$
and the only 
tunable
parameter $r_1$ is related to the $1p$ halo radius.
The mean square root radii for the halo wave states (\ref{ANSATZ-2s}) 
and (\ref{ANSATZ-1p}) are given by
\begin{eqnarray} \label{RADII}
\sqrt{\langle r_{2s}^2\rangle} = r_0
\left( \frac{6(45r_0^4 + 2a^4 - 12 a^2r_0^2   )}
{105r_0^4 + a^4 - 15 a^2r_0^2 } \right)^{1/2}  , \quad 
\sqrt{\langle r_{1p}^2\rangle} = \left(\frac{15}{2}\right)^{1/2}r_1.
\end{eqnarray}

The matrix element of the weak interaction 
(\ref{WEAK-POTENTIAL}),(\ref{W-SP-HALO}) between the ground state
and the first excited state reads 
\begin{eqnarray}   \label{W-MATRIX-ELEMENT}
\langle 2 s | W_{sp} | 1 p \rangle = \qquad \qquad \qquad \qquad
\\
i g^W_n \frac{ G  }{  \sqrt{2} m } 
\int\limits_{0}^{\infty}  dr   
\chi_{2s}(r) \left( \rho_c(r) \frac{d}{d r} + \frac{ \rho_c(r)}{r} +
\frac{1}{2} \frac{d \rho_c(r)}{ d r} \right) \chi_{1p}(r) \nonumber
\end{eqnarray}

The core nucleon density $\rho_c(r)$ has been tuned to reproduce
the data obtained from Ref. \cite{SAGAWA}. We found that their
results are excellently reproduced by the Gaussian-shaped 
ansatz $\rho_c(r)$,
\begin{equation} \label{ANSATZ-CORE}
\rho_c(r) = \rho_0 e^{ -\left(r/ R_c \right)^2}
\end{equation}
with the values of the parameters $\rho_0 = 0.2 fm^{-3}$ and
$R_c = 2 fm$, as shown on Fig.2.

Using 
the model wave
functions (\ref{ANSATZ-2s}),(\ref{ANSATZ-2s}) 
and the core density (\ref{ANSATZ-CORE}),
the required integrals can be done analytically,
and we arrive with the result
\begin{equation} \label{INTEGRAL-ANALYT}
\langle 2 s | W | 1 p \rangle =  i g^W_n \frac{ G }{  \sqrt{2} m }  
{\cal R}
\end{equation}
where 
\begin{eqnarray}  \label{INTEGRAL-ANALYT-R}
{\cal R} = \rho_0 R^3_c C_0 C_1
\biggl\{ 3I_2(y) -\left[ 3\left(\frac{R_c}{a}\right)^2+1\right] 
I_4(y) +
\nonumber\\ +
\left(\frac{R_c}{a}\right)^2 I_6(y)
-
\frac{R_c}{r_1}\left[I_3(y)-\left(\frac{R_c}{a}\right)^2 I_5(y)\right]
\biggr\}
\end{eqnarray}
where $y=\frac{R_c(r_0+r_1)}{r_0 r_1}$ and the functions $I_n$
are given by
\begin{displaymath}
I_n(y) = \int\limits_0^{\infty}dx \quad x^n e^{-x^2-yx} = 
(-1)^n \frac{\sqrt{\pi}}{2}  \frac{d^n}{dy^n} e^{y^2/4} erfc(y/2),
\end{displaymath}
where $erfc(y)$ is the error function 
\begin{displaymath}
erfc(y) = 1-\frac{2}{\sqrt{\pi}} \int\limits_0^y dt \quad exp( -t^2/2).
%
\end{displaymath}
To obtain the results for the PNC weak interaction matrix element, 
we used the parameters $r_0$ and $r_1$ in the halo wave
functions to fit the radial densities of the halos obtained by
Sagawa \cite{SAGAWA}. 

The results for the best parameters
are shown in Figs. 3 and 4 for the $2s$ and the $1p$ halos,
respectively. One sees that the agreement for the densities is 
very good. Below, we use the values 
\begin{eqnarray} \label{VALUES-r0-r1}
r_0 (best \quad value) = 1.45 fm , \qquad 
\nonumber\\
r_1 (best \quad value) = 1.80 fm, 
\end{eqnarray}
to calculate the 
matrix elements in 
Eqs.
(\ref{W-MATRIX-ELEMENT},\ref{INTEGRAL-ANALYT},
\ref{INTEGRAL-ANALYT-R}).
The radial wave functions $\chi$ are given in Fig.5.
We used also deviations of the both $r_0$ and $r_1$ 
from (\ref{VALUES-r0-r1}) to check 
robustness of the results with respect to variations  
in the halo structure details.
The values of the halo radii 
given by (\ref{RADII}),
$\sqrt{\langle r_{2s}^2\rangle}=5.9 fm$ and 
$\sqrt{\langle r_{1p}^2\rangle} = 4.9 fm$ are 
close to 
the values of Ref.\cite{SAGAWA}   $6.5 fm $ and $5.9 fm$
which agree with experimental matter radii.

Substituting the values (\ref{VALUES-r0-r1}) 
into our expressions 
for the matrix elements we obtain the following value of the 
matrix element $\langle 2 s | W_{sp} | 1 p \rangle_{HALO}$
\begin{eqnarray} \label{W-RESULT-NUMERIC} 
\langle 1 p | W_{sp} | 2 s \rangle_{HALO} = - i 0.2 
\quad  g^W_n \quad eV,
\nonumber\\
= - i 0.2   \quad eV \quad 
( for  \quad g^W_n \simeq 1 \quad).
\end{eqnarray}
It is seen that this value is only few times smaller than the 
standard value of the matrix element of the weak potential 
between the opposite parity states in spherical nuclei 
(see e.g., \cite{ADELBERGER}), that is typically about one $eV$.
This results from the wave function structure 
and comes basically from the facts that the $2s$ wave function crosses
zero line near the core surface while the $1p$ radial wave function 
does not have nodes. Thus the functions $\chi_{1p}$ and 
$d \chi_{2s}/ dr $ 
look similar and are folded constructively with $\rho_c(r)$ in the
region of interaction (nuclear interior), see Fig.6.

The matrix element of $W_{sp}$ between the ``normal'' 
nuclear states can be evaluated for example, in the oscillator 
model. Taking the typical matrix element between the states 
$2s$ and $1p$ and using the same formula (\ref{W-MATRIX-ELEMENT}), 
one has
\begin{equation}  \label{W-MATRIX-OSC}
\langle 1 p | W_{sp} | 2 s \rangle_{osc} = 
- i  g^W_n  G \rho_0   
\left(\frac{\omega}{2 m}\right)^{1/2}
\end{equation}
where $\omega \simeq 40 A^{-1/3} MeV$ is the oscillator frequency
\cite{BM} with $A$ the nuclear mass number.
We used here the constant value of the core nucleon density, 
$\rho_0 \simeq  0.138 fm^{-1/3}$. 
This is very good approximation in the case of normal 
nucleus \cite{we}.
 
Recalling 
the energy difference between the ground state 
and the first excited state $1p$ that is known experimentally,
\begin{equation} \label{DELTA-E}
| \Delta E_{HALO} | = E_{p 1/2} - E_{s 1/2} = 0.32 MeV 
\end{equation}
we obtain, using Eq.(\ref{W-RESULT-NUMERIC}), 
the coefficient of mixing the opposite parity state ($1p$)
to the halo ground state $2s$:
\begin{eqnarray} \label{HALO-MIXING}
|f^{HALO}_{sp}| = \frac{ | \langle 1 p | W_{sp} | 2 s \rangle | }
{ |\Delta E_{HALO}| }
\quad \simeq \frac{0.2 eV g^W}{ 0.32 MeV} 
\nonumber\\
\simeq 0.6 \times 10^{-6} g^W_n
\nonumber\\
\simeq  0.6 \times 10^{-6} \quad (for \quad g^W_n \simeq 1)
\end{eqnarray}
This PNC mixing is about one order of magnitude stronger than the
scale of single-particle PNC mixing in ``normal'' nuclear states
that can be extracted from Eq.(\ref{W-MATRIX-OSC}). In the case of 
normal $p-s$ mixing, we have 
\begin{eqnarray}  \label{NORMAL-MIXING}
|f^{normal}_{sp}| = \frac{ | \langle 1 p | W_{sp} 
| 2 s \rangle | }{ \omega }
= \frac{ G g^W_n \rho_0 }{ \sqrt{2} m } 
\left( \frac{m }{\omega} \right)^{1/2} 
\nonumber\\
\simeq 0.7 \times 10^{-7} g^W_n
\nonumber\\
\simeq 0.7 \times 10^{-7} \quad (for \quad g^W_n \simeq 1)
\end{eqnarray}
in the same region of nuclei with $A \sim 11$. 
The above value (\ref{NORMAL-MIXING}) for the normal PNC mixing is
rather universal and it is  practically insensitive 
to variations of the details of the normal nuclear wave functions
and core densities \cite{we}. 
One should stress that
in the halo case, the energy denominator in 
Eq.(\ref{HALO-MIXING})
is $\omega/| \Delta E_{HALO} | \simeq 50$ times
smaller than in the normal case 
(\ref{NORMAL-MIXING})
based on  the oscillator model.
%
Comparing Eqs.(\ref{HALO-MIXING}) and
(\ref{NORMAL-MIXING}), we find the halo enhancement factor 
in the PNC mixing
to be
\begin{eqnarray}  \label{HALO-ENHANCEMENT}
\frac{|f^{HALO}_{sp}|}{|f^{normal}_{sp}|} \quad \simeq \quad 9 .
\end{eqnarray}
This result is quite remarkable in a number of respects.
First, it is seen that 
in experiments when the halo wave functions in nuclear exterior 
are probed,
the value of PNC mixing is even stronger than in ``normal'' nuclei.
Secondly, this PNC mixing is dominated by the neutron weak constant 
$g^W_n$. Such experiments with neutron halo nuclei
would therefore provide unique opportunity
to probe the value of this constant.
Usually, sensitivity of experiments 
to the value of this constant is ``spoiled'' by comparably large 
value of the proton weak constant $g^W_p$, cf. Eqs.
(\ref{g-values}).

In order to assess reliability of the results,
we have studied stability of the enhancement factor 
against variations in the parameters of the halo wave functions.
As one can see from the results presented in Table I,
the matrix element (\ref{W-RESULT-NUMERIC}) is changed by few 
per cent only when the wave functions are deformed. The enhancement 
factor (\ref{HALO-MIXING}) is therefore quite stable. 

\section{Conclusion}
\label{sec:level6}

Having in mind to present a first estimate of the PNC effect in
halo nuclei, we have chosen here the simplest possible case
of one-body halo 
where the existing data
allow one to rely on simple analytical model
of halo structure.
In this work, we confined ourselves to the case of exotic nucleus 
$^{11}Be$ for which we presented detailed consideration. 

The analysis presented above rests basically on the most reliably 
known facts: the quantum numbers of the states involved, the halo
radii which match the matter radii known from experiment, 
and the Hartree-Fock wave
functions. With these input data, the further quantitative analysis 
is a straightforward analytical exercise which does not require 
any approximations. Stability of the results has been checked 
analytically. The 
PNC enhancement factor 
of one order of
magnitude allows one to to speak about qualitative halo effect
that should not be overlooked.

It is the matter of further studies to check universality of
the effect while going along the table of exotic nuclei. 
One sees that other exotic nuclei with developed 
halo structure manifest similar properties (see, e.g., \cite{SAGAWA}).
Indeed, the effect of PNC enhancement found here 
results basically from the two facts:

(i) small energy separation between the mixed opposite parity states

(ii) considerably strong overlap between the mixed wave functions and
the core density, which saves part of suppression in the PNC weak 
matrix element.

The first of these points is rather common for nuclei with 
developed neutron halos. Systematics of separation energies for 
single neutron \cite{HALO-OBSOR} shows that the ground  
states of halo nuclei can be distanced from the continuum by 
typical spacing 
$\varepsilon_{halo} \sim ( 2 m r_{halo})^{-1/2} \sim $ 
few hundreds of KeV. Even in the cases when no bound states with
parity opposite to that of the ground state occur, the PNC
admixtures to the ground state wave functions  must exist. 
In these cases, the PNC admixtures can be evaluated 
by means of Green function method.

The second point (ii) is 
related to the wave function structure and requires further studies.
It would be also interesting to study the PNC effects in proton rich
nuclei \cite{Brown-Hansen}, \cite{Karataglidis-Bennhold}, 
\cite{Ren-Faessler-Bobyk}.

One should mention that the results obtained here are based on
the single-particle approximation (one-body halo model).
In principle, the halo neutron can couple to excitations of the core
(see, e.g. \cite{CORE-EXC}). In fact, such coupling can be 
responsible for  
the small energy separation between the opposite parity levels 
in $^{11}Be$, which can be separated by few MeV otherwise 
(see, e.g., \cite{SY}).  
These many-body effects may be important for precise 
evaluation of the PNC mixing, which is beyond the scope of this paper.
We did not consider contribution from such
effects here.
 
In conclusion, we have shown in this work that the ground state 
of exotic nuclei with developed halo structure contain mixing of
parity violating admixtures of opposite parities, using the nucleus 
$^{11}Be$ as a representative case. 
Originating
at the core scale, where the halo wave functions 
overlap strongly with the core nucleon wave functions, these
PNC admixtures can manifest themselves in the nuclear exterior,
where the halo neutron wave functions contain admixtures  
that are about one order of magnitude bigger than in the case 
of normal nuclear states. Such PNC admixtures can be tested 
in experiments that can probe halo 
wave functions in nuclear exterior.
Moreover, it is important that 
the PNC admixtures in neutron halo
are sensitive to the neutron weak 
constant $g^W_n$, thus providing a new 
interesting possibility to probe
this weak constant whose value is the most doubtful point of the
present theory of PNC weak interaction.

One of possible experimental manifestations of the discussed
effect is related to anapole moment \cite{ZELD},\cite{FKh}
which attracts much attention in current literature \cite{Haxton-SC}
in view of new experimental results (detection of anapole moment
in nucleus $^{133}Cs$ \cite{EXP-AM-SC}).
Since the anapole moment is created by the toroidal 
electromagnetic currents 
which results from PNC, its value grows as the size of 
the system increased \cite{FKh}. 
In the case of halo which we considered here, the value of the 
anapole moment can be therefore enhanced due to extended halo cloud.
We hope to address these issues in following publications.

\section{Acknowledgement}
\label{sec:nolevel}
The authors are grateful to V.V.Flambaum for discussions of
of issues related to PNC effects.
The work has been supported by FAPESP 
(Fundacao de Amparo a Pesquisa do Estado de Sao Paulo) and  
in part by funds
provided by  the U.S.~Department of Energy (D.O.E.) under contract
\#DE-FC02-94ER40818.

\newpage
\begin{table}
\caption{Stability analysis for the matrix element of $W_{sp}$ 
between 
the halo states $2s_{1/2}$ and $1p_{1/2}$. 
The results for the values
of the parameters $r_0$ and $r_1$ differing from the 
best values are
shown. The central entry in the table corresponds to the best 
value.
It is seen that variations in $r_0$ and $r_1$ do not affect 
$\langle 2s_{1/2} |W_{sp} | 1p_{1/2} \rangle$ any considerably.
} 

\vspace{1cm}

\begin{tabular}{cccc}   
       & $ r_0=1.40 $ & $r_0=1.45$  & $ r_0=1.50$  \\ 
\tableline
$ r_1= 1.75 $   &  $1.168 $  & $1.052 $  &  $0.950$  \\
 $ r_1= 1.80   $ &  $1.110 $  & $1.000 $  &  $0.903 $  \\ 
  $ r_1=  1.85   $ &  $1.056 $  & $0.952 $  &  $0.860 $  
\end{tabular}

\end{table}

\newpage

{\large Figure Captions}

\vspace{1cm}
Fig.1. 
a) Schematic plot of matter distribution in halo nuclei.
The dark region corresponds to the nuclear core, the grey  
region shows the halo neutron cloud. 

b) The spectrum of the bound states $^{11}Be$.

c) Illustration of the single-particle PNC mixing in the ground
state of $^{11}Be$.

\vspace{1cm}

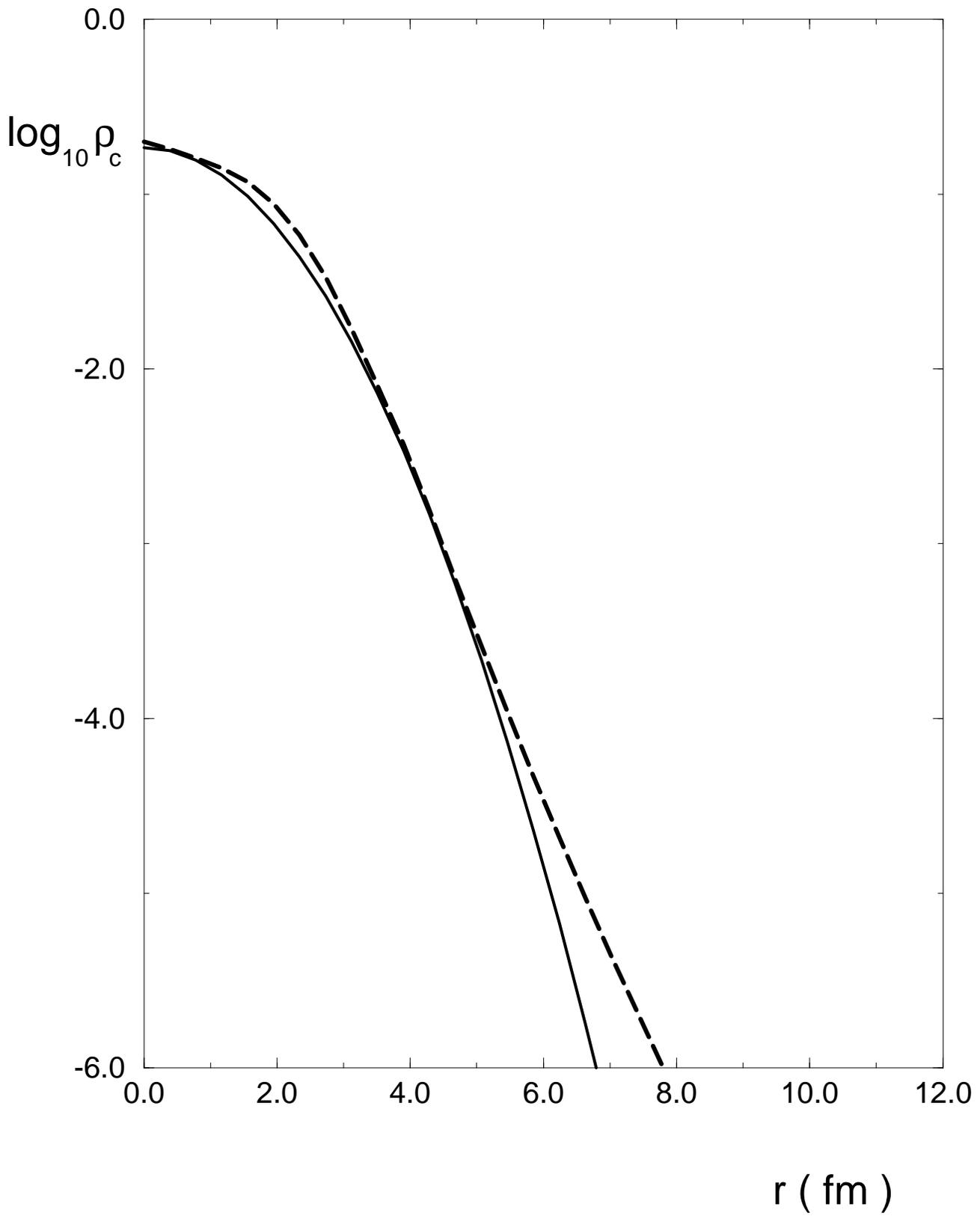
Fig.2. The core density distribution (logarithmic scale).
The dashed line corresponds to Ref. \cite{SAGAWA}, the solid line 
gives parametrization (\ref{ANSATZ-CORE}).

\vspace{1cm}
Fig3. The halo density in the ground state,
$\rho_{2s1/2}(r) = \frac{1}{4 \pi} \left( R_{2s1/2}(r) \right)^2$.
The dashed line corresponds to the Hartree-Fock calculations of
Ref. \cite{SAGAWA}, the solid line 
gives parametrization (\ref{ANSATZ-2s}),(\ref{VALUES-r0-r1}).

\vspace{1cm}
Fig.4. The halo density in the first excited state,
$\rho_{1p1/2}(r) = \frac{1}{4 \pi} \left( R_{1p1/2}(r) \right)^2$.
The dashed line corresponds to the Hartree-Fock calculations of
Ref. \cite{SAGAWA}, the solid line 
gives parametrization (\ref{ANSATZ-1p}),(\ref{VALUES-r0-r1}).

\vspace{1cm}
Fig.5. Plot of the radial wave functions 
of the states $|2s1/2\rangle$
and $|1p1/2\rangle$,
$\chi_{2s1/2}(r) = r R_{2s1/2}(r)$ and $\chi_{1p1/2}(r) = 
r R_{1p1/2}(r)$. 

\vspace{1cm}
Fig.6 .
Plot of the functions contributing to the 
weak PNC matrix element.
The function $s(r)=\frac{d}{d r} \chi_{1p1/2}(r) + 
\frac{\chi_{1p1/2}(r)}{r}
+\frac{d\rho_c/dr}{2\rho_c}\chi_{1p1/2}(r)$ (dot-dashed line) 
depends on $r$
in the way similar to $\chi_{2s1/2}(r)$ (dashed line).
The combination $\chi_{2s1/2}(r) \rho_cs(r)$ that enters
the PNC matrix element in Eq.(\ref{W-MATRIX-ELEMENT}) 
is shown by the solid line. It contributes coherently to
$\langle 2 s | W_{sp} | 1 p \rangle$.

\noindent
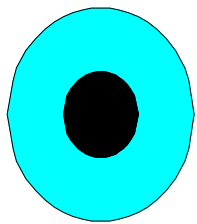
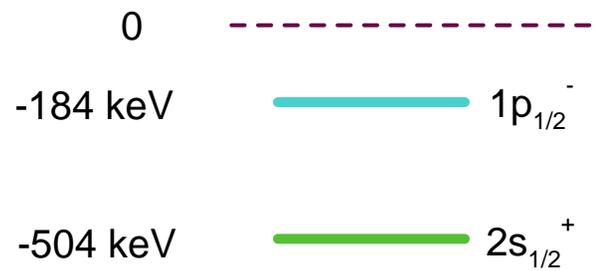
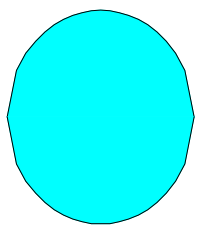
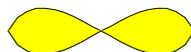
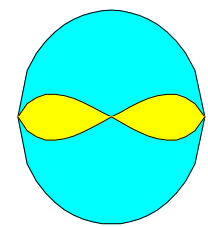
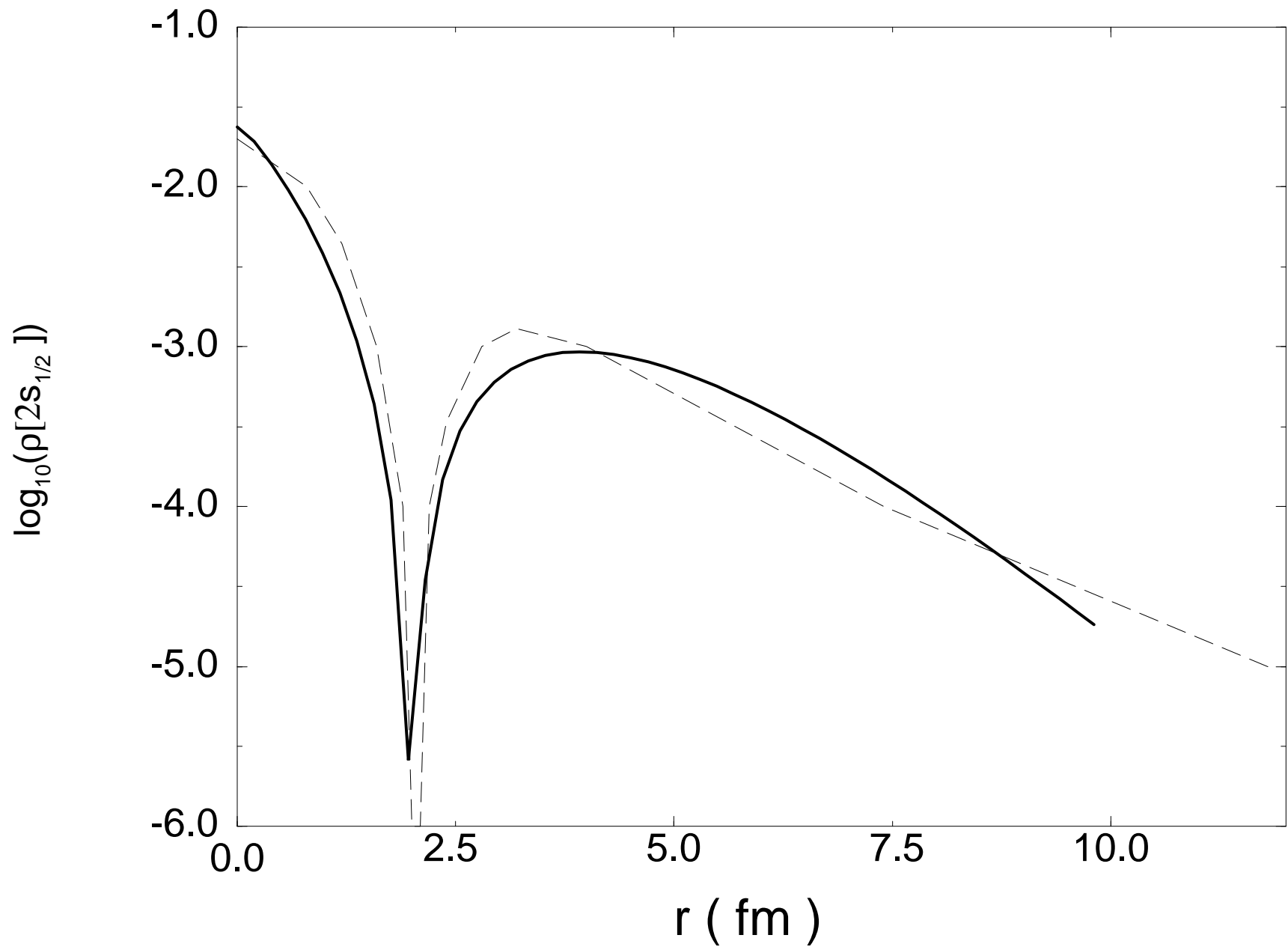
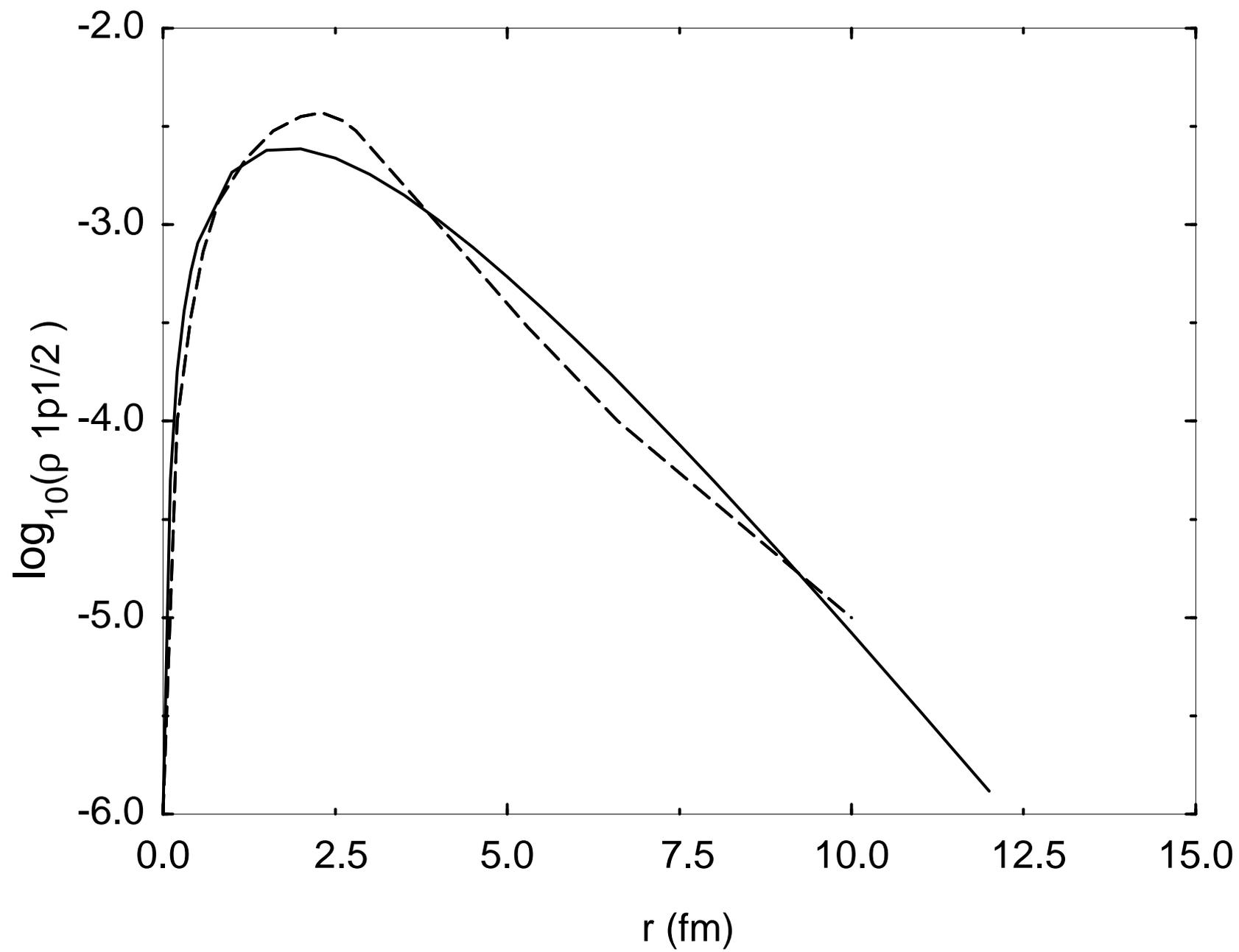
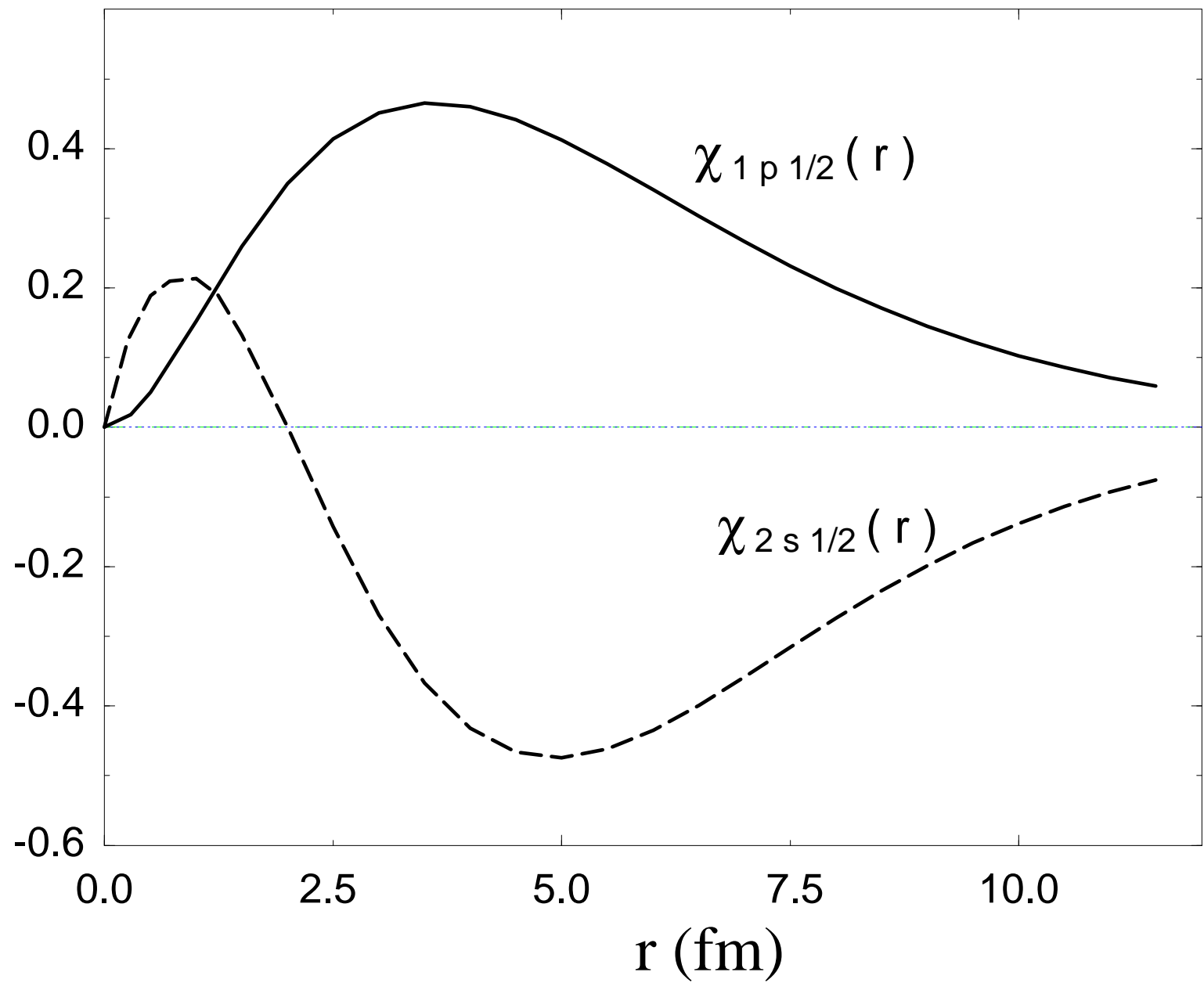
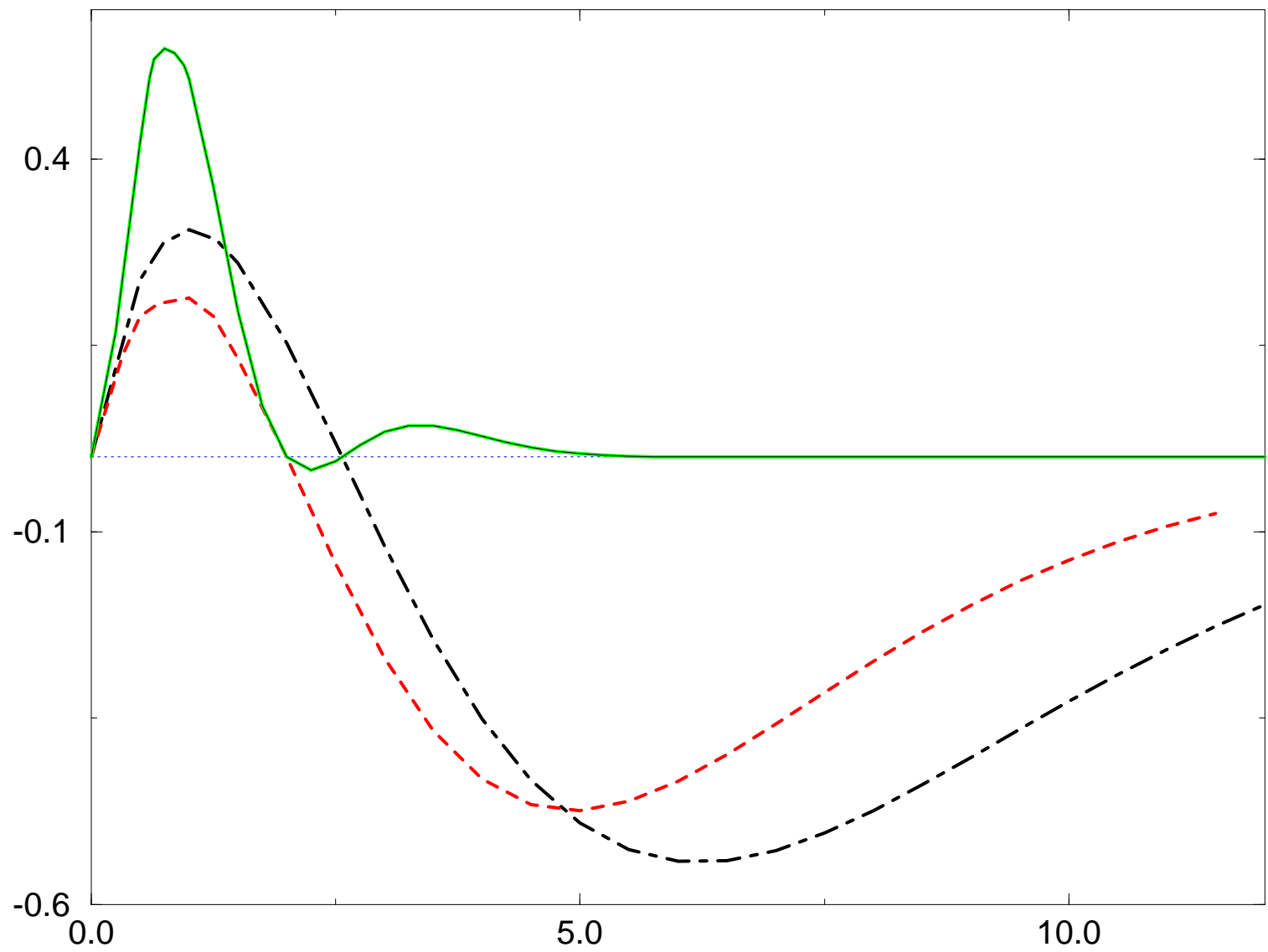
\end{document}